# A Short Introduction to Basic Principles of Quantum Navigation Based-on Rb Cold Atom Interferometry

Narges Kafaei [1][*] and Ali Motazedifard [2][*]

[1] *Institute of Quantum Sciences and Technologies, University of Tehran, Tehran, 143995596, Iran*
[2] *Quantum Sensing & Metrology Group, Iran Center for Quantum Technologies (ICQT), Tehran 15998-14713, Iran*
[*]Corresponding authors E-mail: n.kafaei@gmail.com & Motazedifard.ali@gmail.com

**Abstract:** *Due to advances in cold atom interferometry, laser spectroscopy it is possible to achieve more precise accelerometers and gyroscopes which never occurs in mechanical- and optical-based sensors. Because of the inherent and independent characteristics of atomic levels which are too sensitive respect to the external changes, a self-calibrated navigation system with no satellite can be achieved. Here, in this paper we very shortly review the basic principles of inertia cold atom navigation sensor.*

**1. Introduction**

Each navigation system includes a collection of satellites and ground stations that can provide a variety of positioning, navigation, and time services at the public and military levels. The most important key elements in navigation systems are gyroscopes and accelerometers. The main challenge is the limited accuracy of gyroscopes [1]. As can be seen in Fig. 1, the best mechanical gyroscope is about $10^{-6°}/h$, while the best optical gyroscope has an accuracy of about $10^{-4°}/h$ [2].

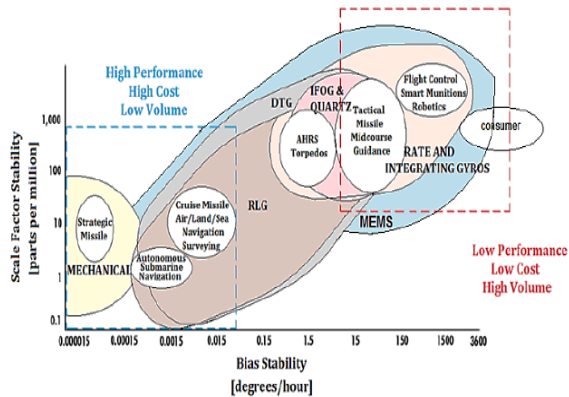

Fig. 1: Comparison between optical and mechanical gyroscopes [2].

With recent advances in atomic interferometry with ultra-cold coherent atoms ultra-precise accelerometers are achievable. In fact, by creating interference between the ground and excited states of the cold atoms via laser pulses locked on the permissible levels, one can see the phase differences due to acceleration or external factors in the interference spectrum of atoms or intensity changes. This enables us to ultra-precise measurement. The accuracy of the cold-atom interference gyroscope based on the Sagnac effect as representative of the latest generation gyroscope is reported about $10^{-12°}/h$ [3]. The most important feature of atomic navigation is its self-calibration that requires no satellite station information for calibration and positioning.

**2. Physical Principles of Laser Cooling**

To emerge the full quantum coherence signature in atomic ensembles one should remove the thermal effects by cooling the atoms. More technically, the atomic wave function indicating probability density must be localized for a sufficient period at the volume required to perform the interaction between light and matter. This is usually achieved by trapping atoms into a magnetic optical trap (MOT). MOT is composed of combining the magnetic field with three pairs of laser beams emitted in opposite perpendicular directions that interact with the atom in a way that gradually slows it down at the center. The magneto optical trap (see Fig. 2) includes a weak position-dependent quadrupole magnetic field coils (Helmholtz coil) and three pairs of orthogonal red-detuned circular polarization optical beams for cooling.

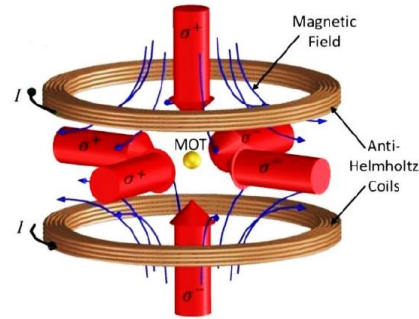

Fig. 2: Magneto optical trap arrangement.

When the atoms move away from the zero field in the center of the trap (halfway between the magnetic field-producing coils), the Zeeman shift causes the laser light to move from the opposite direction, despite being away from the resonance (red detuning) to the resonance. Therefore, atoms sense pure force from the beam at the resonance toward the center. This force arises from photon scattering, in which atoms receive impulse momentum in the opposite direction of their motion, and thus cooled down by frequent spontaneous absorption and emission. In this way, a magnetic-optical trap can trap and cool atoms at initial speeds of hundreds of meters per second to tens of centimeters per second. Only certain atoms and ions have optical transitions that are compatible with laser cooling, because the large laser power required at wavelengths much shorter than 300 nm is very difficult. In addition, an atom with more hyperfine structures has more selections to emit the photon from the upper state and not return to its original state which leads the atom to be in a dark state and destroys the cooling process. Other lasers can be used to pump the atoms into an excited state and try again, but the more complex hyperfine structure needs more lasers (narrow-band, frequency locked). This additional repumping narrow-band and stable lasers dictates more complexity and prices. That is why the $^{87}$Rb is a very good candidate for laser cooling, as it can be cooled with an inexpensive diode laser, for example using typical narrow-band ECDL or DBR lasers.

## 3. Vacuum System

The MOTs usually must be loaded in a ultra-vacuum chamber. Because the potential to be trapped in a magnetic optical trap is small compared to the thermal energies of atoms, most collisions between trapped atoms and the underlying gas provide enough energy to escape the trapped atom. If the background pressure is too high, the atoms will be removed from the trap faster than they can be loaded, and traps will not form. This means that the optical magnetic trap cloud forms only in a vacuum chamber with a background pressure less than $10\mu$Pa.

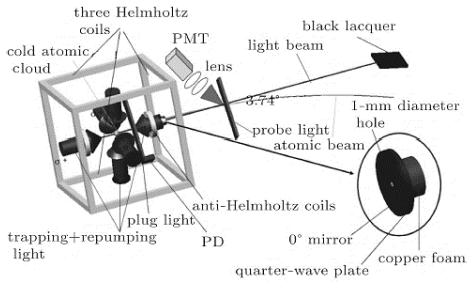

Fig. 3: MOT configuration [4].

Since ultra-high vacuum is required for the proper operation of trap, the whole system needs to be completely removed from the gas at high temperatures by long-term pumping with a molecular turbo pump, and later with an ion pump. The cell is then placed in the center of the magnetic coil system. In this experiment reported in Ref. [4], a two- vacuum chamber system consisting of a 3D MOT system and a probe system for atomic beam is required. The two chambers are connected by a 1mm hole. In the MOT chamber (Fig. 3), an unbalanced trap is produced and an atomic beam is achieved from a trapped cold atomic cloud. In the probe chamber, the atomic beam is monitored via the fluorescence spectrum and time of flight. The rubidium chamber is connected to the MOT through a tap. Two ion pumps are needed to put the two chambers under background pressure $8\times 10^{-10}$ torr and $2\times 10^{-9}$ torr (when the rubidium chamber is turned off). When the rubidium chamber is turned on and heated to about 36 to 40°C, the rubidium vapor must be entered into the MOT chamber. In this case, the pressure of the chamber will increase almost $10^{-7}$ torr immediately, but the pressure of the chamber of the probe will not change significantly. Vacuum chamber windows should be coated with broadband infrared anti-reflection (NIR) coatings and connected to the enclosure using indium wire for sealing. At the top of the chamber is a 0° mirror and a λ/4 wave plate with a diameter of 2 inches with a 1mm hole in the center. A tube of copper foam must be inserted into the probe chamber just after the hole to absorb the remaining atoms leaking from the MOT chamber.

## 4. Laser System

To perform $^{87}$Rb laser cooling, two optical frequencies are required. As shown in Fig. 4, the first laser, set around the transition cycle $|5S_{1/2}, F=2\rangle \rightarrow |5P_{3/2}, F'=3\rangle$, is used for the cooling process. The second laser is set as a repumping laser on transition $|5S_{1/2}, F=1\rangle \rightarrow |5P_{3/2}, F'=2\rangle$, and removes the atoms from the ground state back to the cooling cycle. The difference in frequency between these two transitions is 6.568GHz, which is provided by an electro-optic or acousto-optic modulator and transmission from the main beam. Six laser beams that are mutually perpendicular to each other are focused on the zero center of the quadrupole magnetic field for cooling purposes. Two fiber laser sources at 780nm wavelength are essential for producing laser repumping and cooling frequencies for laser cooling. To this end, two ECDLs (Fig. 5) that radiate 1560nm wavelength can be used, which after amplification with two Yb-Er crystal amplifiers, their frequency is doubled by nonlinear PPLN crystal and converted to 780nm. The linewidth of ECDLs is narrow (5kHz) and can be adjusted to about 10GHz. Diodes are set in the room temperature range and operate at a current of about 90mA to deliver a 15mW optical output power. Frequency control is achieved by applying Feedback to the current of the diode. The LD1 laser is used for the repumping transition. Its frequency is locked by saturation absorption spectroscopy (SAS) on the dominant cross-over signal between the atomic states $|5P_{3/2}, F'=1\rangle$ and $|5P_{3/2}, F'=2\rangle$, and its red shift with the repumping transition frequency is 78.5MHZ.

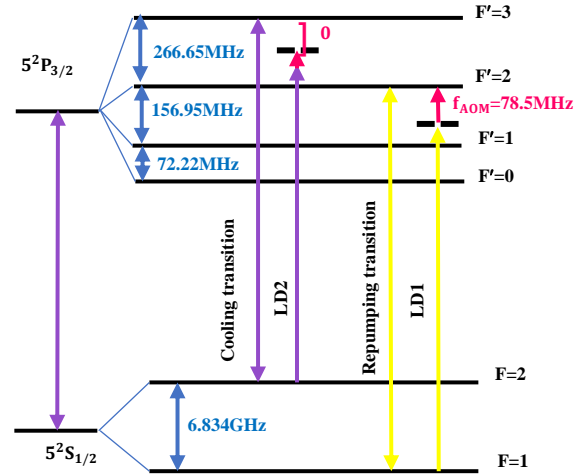

Fig. 4: Hyperfine structure of D2 transition of $^{87}$Rb.

This laser is then used as an absolute frequency reference to control the LD2 frequency, which creates the cooling transition. The difference in frequency between LD1 and LD2 in 1560nm is set at about $v_{12} = 3.284\,\text{GHz}$.

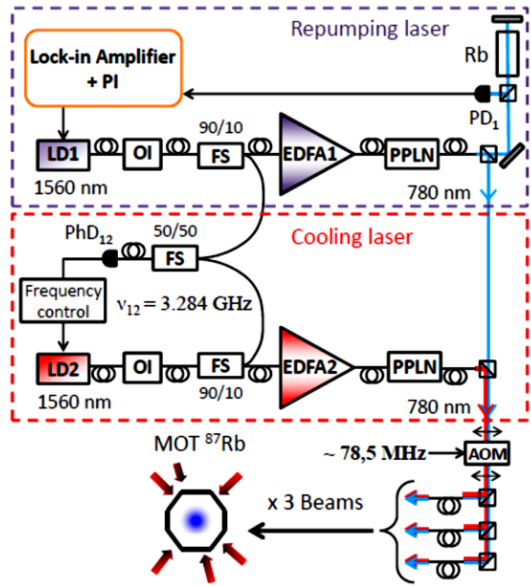

Fig. 5: Laser system for laser cooling of $^{87}$Rb [5].

In order to provide sufficient optical power, the LD1 and LD2 output signals are injected into two EDFA fiber amplifiers. In normal operation, ~150mW-LD1 and ~360mW-LD2 is achieved after amplification. After doubling the frequency, the frequency difference between the two lasers is 6.568GHz, which corresponds to the frequency difference between the repumping transition and the cooling. In this configuration, ~45mW and ~180mW power for repumping and cooling lights at the output of the frequency doubling phases are achieved. The output beams are then combined and the frequency is changed by an acosto-optic modulator (AOM). The AOM is fed with frequency $f_{AOM} = 78.5\,\text{MHz}$ to fill the frequency gap between the repumping transition and the cross-over signal that LD1 is locked on. This arrangement enables the laser to control the intensity of the light by changing the input RF power to AOM. The three output fiber beams, containing about 30mW laser cooling and approximately 4mW repumping lasers, can be used to create a magnetic optical trap in a reflective configuration. Finally, trapping and repumping light are coupled using a beam splitter (BS) and the fiber coupling injected into the amplifier and divided into three sections by a fiber splitter and directed into the MOT chamber. A fiber system is needed to transfer light from the optical table to the vacuum system. All fibers must maintain polarization and be attached to collimators to remain fixed on the chambers. Each collimator consists of a λ/2 half wave plate, a polarized beam splitter (PBS), a λ/4 quarter wave plate and a 200mm focal lens collimator. Through collimators, three parallel beams can be received with adjustable intensities. These beams must be reflected in opposite directions by three sets of reflective optics that produce six cross-laser beams in the MOT chamber to have 3D-MOT. The image of the atomic beam can be viewed by a CCD camera through a window in the MOT vacuum to sure its existence. The frequency of the LD1 laser is doubled to amplify the cooling process by repumping. This laser is locked on an $^{87}$Rb saturation absorption signal using frequency and amplitude modulation spectroscopy [6]. The current of the laser diode is modulated at a frequency of 100kHz. After doubling phase, the laser light is sent at 780 nm to the rubidium cell to achieve saturation absorption spectroscopy.

In the saturation absorption spectroscopy according to Fig. 6, two laser beams are used as pump (strong) and probe (weak) in opposite directions. The atom moves towards whichever of these beams it is resonated with that beam due to Doppler shift. When the atom's velocity is zero, both resonant beams occur, but because in this case the pump is saturated, the absorption of the probe is negligible. Therefore, in the absorption peak, a peak caused by saturation. Since the probe signal is detected, a strong signal is detected in the photodiode because there is no absorption. The saturation spectroscopy output signal is collected by a photodiode (PD1, with a bandwidth of 10MHz).

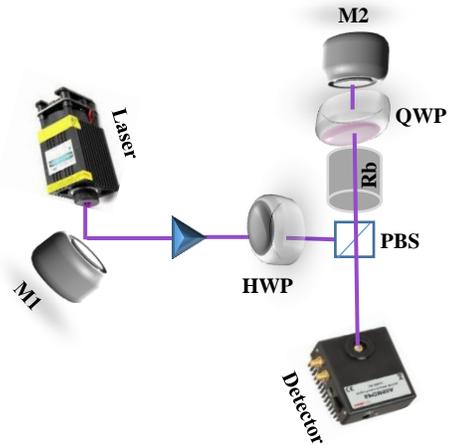

Fig. 6: Schematic setup of saturation absorption spectroscopy.

The output signal from the photodiode, in addition to the modulation frequency (100kHz), has other harmonics. These harmonics that have absorption effects are detected by a mode locking device. An error signal is produced to correct signal from Doppler-free SAS and applied to the diode injectable current to correct and stabilize the wavelength of the diode laser. The laser diode temperature is set at 31°C through internal temperature control. In SAS, lock-in amplifier is used. Locking the laser frequency on the resonance frequency of rubidium requires the creation of a feedback loop, in which an error signal, proportional to the laser decoupling of the desired frequency, is returned to the driver laser controller to make the necessary adjustments. Using an amplifier, the signal $signal = noise + s_1 \sin(\nu t) + s_2 \sin(2\nu t)$ is taken from the photodiode and multiplied by a sine wave at the reference frequency $\nu$. By integrating it into $\lim_{T\to\infty} 1/T \int_0^T signal \cdot \sin(\nu t) dt = 1/2 s_1$, the required error signal $s_1$ is transmitted to the laser controller to adjust the corrections needed to stabilize the laser frequency. The LD2 laser is used for cooling transition $|5S_{1/2}, F=2\rangle \to |5P_{3/2}, F'=3\rangle$ after doubling frequency. To perform sub-Doppler cooling, the red detuning between the laser frequency and the cooling transition must be adjustable between 0 and $-20\Gamma$, where the $\Gamma = 6.06$ MHz is the natural linewidth of the D2-line is $^{87}$Rb. To meet these requirements, LD1 is used as a frequency reference. As mentioned, the LD1 laser frequency is stabilized by SAS with amplitude and frequency modulation, but LD2 laser frequency verification has difficulties. Therefore, the difference in frequency between LD2 and LD1 is locked and controlled by a microwave system according to Fig. 7. Small amounts of light (10%) of LD1 and LD2 are placed on a fast photoconductor (PhD12, bandwidth: 15GHz). In order for the photoconductor to be fast, the bandwidth must be large. Then, according to Fig. 5, the beat note frequency at $\nu_{12} = 3.284$ GHz is produced, amplified and combined with a reference signal. This reference signal was generated by a microwave frequency synthesizer at 1.562GHz and has doubled its frequency. Again, these frequencies are performed beat note, and their frequency difference of $\nu_{int} = 160$ MHz is generated. Therefore, the difference between the frequency of two lasers varies from GHz to MHz.

$\nu_{vco} = 152$ MHz and a filter that can pass below 100MHz, the frequencies 160MHz and 152MHz will not pass through the filter, but the difference of 8MHz will pass through the filter. Using a splitter, the resulting signal frequency at 8MHz is divided by 16 and achieved at 500kHz, which is comparable to the frequencies of saturation spectroscopic wave modulation. The output signal is sent to the frequency to voltage converter (FVC), which creates an error signal corresponding to the frequency difference between the two lasers. This signal is compensated by $V_{offset}$ to set the signal to zero when the frequency difference between $\nu_{vco}$ and $\nu_{int}$ is 8MHz. After shaping and filtering, this output signal is used to generate feedback, which allows the frequency lock of the LD2 laser on LD1. If the frequency of the LD2 laser moves, the frequency of the LD2 laser will move as well as 500kHz, because the two lasers will no longer be 3.284GHz. Since the frequency of the second laser has changed, the frequency difference between the two lasers will also change. With the help of an integral circuit, this change is detected and a signal is sent to the current of the second diode laser to stabilize its frequency until it returns to 500kHz. So, the LD2 frequency is stabilized by using the stabilized LD1 frequency.

## 5. Atomic Interferometer

The output atomic beam of the 3D MOT is at the ground state $F = 2$. Before entering the interferometer circle, this beam must be in state $F = 1, m_F = 0$. Therefore, the atomic beam of MOT in the ground state $F = 2$ is pumped in a single hyperfine state $F = 1$ by the laser pumping 1 set to the transition $F = 2 \to F' = 2$ (Fig. 8)) [7]. In this case, atoms are distributed, asymmetrically, in all hyperfine levels of the ground state where the Raman transition spectra is excited by the pulse $\pi$. Another linear polarized optical pumping beam, laser 2, which is set on $F = 1 \to F' = 0$, is used to pump most atoms to $F = 1, m_F = 0$ from other hyperfine levels with $\Delta m = 0, \pm 1$. In these settings, the transition value $\pi$ of the $F = 1 \to F' = 0$ must be minimized, otherwise atoms are dispersed at the level $F = 1, m_F = 0$ to other hyperfine levels of the magnetic field-sensitive ground state.

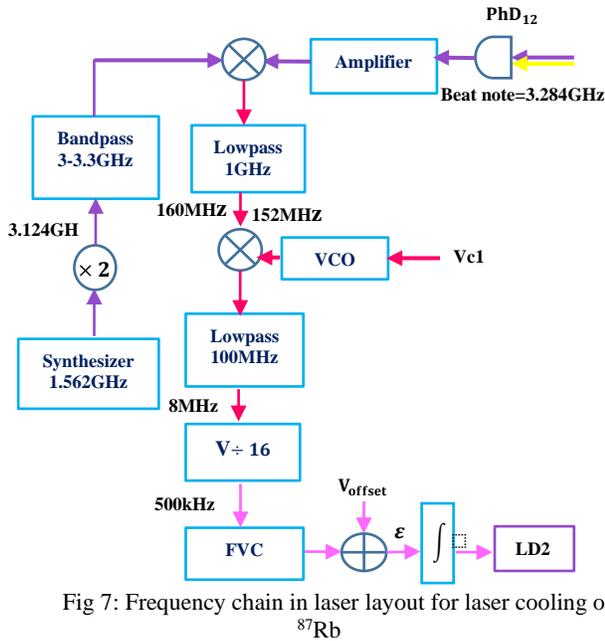

Fig 7: Frequency chain in laser layout for laser cooling of $^{87}$Rb

If the output signal of 160MHz is again combined with the signal given by a voltage control oscillator (VCO) at

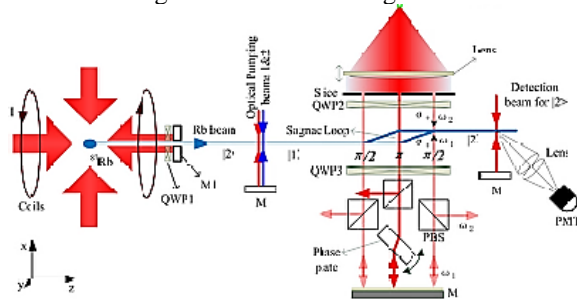

Fig. 8: Experimental setup of atomic interferometry system based on cooled atomic beams of Rubidium [7].

To manipulate atomic wave packets, continuous Raman beams with constant pulse width $d = 1\,\text{mm}$ are used while they are in long-lived ground states. The optical system of Raman lasers is shown in Fig. 9.

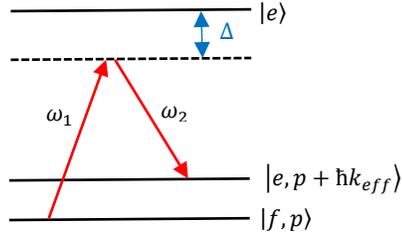

Fig. 9: Raman transitions.

Figure 10 shows the atomic states of an $^{87}$Rb atom that are placed in the path of interferometry. The optical transition between the two states $|f,p\rangle$ and $|e, p+\hbar k_{eff}\rangle$ takes place using the Raman effect induced by two propagated laser beams in opposite directions with frequencies $\omega_1$ and $\omega_2$. The Raman pulse duration is constant and equal to $\tau = d/v_z$, where $v_z$ represents the atomic longitudinal velocity. The Rabi frequency of $\Omega$ is adjusted by changing the intensity of the $\omega_1$ and $\omega_2$ Raman lasers, so the Rabi phase of the $\phi = \Omega\tau$ can be set $\pi$ or $\pi/2$. The amplitudes of the excited spectra are adjusted by two $\pi/2$ pulses equally, and half the amplitude of the $\pi$ pulse.

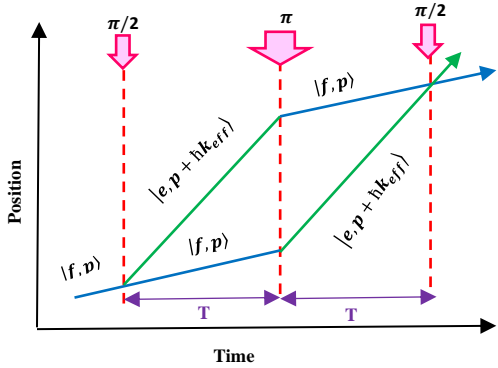

Fig. 10: Atom state in a three-pulse Mach-Zehnder interferometer.

The momentum of two atomic states differs as $\hbar k_{eff} = \hbar(k_1 - k_2) \cong 2\hbar k_1$, where $k_1$ and $k_2$ are the wave vectors of laser beams. Since this method uses a two-photon transition 'a three-pulse sequence $\pi/2 - \pi - \pi/2$ is sufficient to cancel the transverse velocity dependence. The first atomic transition pulse creates a 50% probability from the ground state $|f,p\rangle$ (which was in that state before the atom interferometer was entered) into the excited state $|e, p+hk_{eff}\rangle$. After such a pulse, the atom is in a superposition of two states, meaning that it has remained in the ground state at the same time and continued its initial path in the blue path in Fig. 10, and has been excited by the impulse of the $hk_{eff}$, changing its course to the green line. In other words, for an atom initially in state $|f,p\rangle$, superposition between states $|f,p\rangle$ and $|e, p+hk_{eff}\rangle$ takes place as $(|f,p\rangle - ie^{-i\phi_1}|e, p+\hbar k_{eff}\rangle)/\sqrt{2}$. After time T, the pulse $\pi$ shines on the atom, which causes atomic transition with a 100% probability. Thus, the excited state (where the probability of the atom is 50%) becomes the ground state and the ground state (in which the atom has a 50% probability of being present) becomes the excitation state. In other words, the second pulse changes the internal state and the momentum of two paths as $|f,p\rangle \to -ie^{-i\phi_2}|e, p+\hbar k_{eff}\rangle$ and $|e, p+\hbar k_{eff}\rangle \to ie^{i\phi_2}|f,p\rangle$. Again after the T interval the atom undergoes a second pulse $\pi/2$. Such a pulse with a 50% probability transforms the ground state (in which the atom is present with a 50% probability) to the excitation state and simultaneously with the probability of a 50% excited state (in which the atom is present with a 50% probability) to the ground state. After this pulse, the atomic wave function is reported with the sum of two terms that have the same weight and are associated with the atom in the ground state and in the excited state, respectively. These two terms have different phases because each time a laser pulse has interacted with the atom in sequence, the phase change equals the optical phase in time and space interacts with the atomic wave function, whose signal depends on the atomic transition.

By tilting the Raman beams slightly away from the perpendicular direction with the atomic beam, the Doppler-sensitive transition peaks are precisely displaced from the remaining Doppler insensitive transition peaks as a result of impure polarization of Raman lasers. Doppler displacements and recoil displacements are compensated by compensation for the difference in the frequency of $\omega_1 - \omega_2$. The divergence of the cold atomic beam is slightly higher than the fast thermal beam at the same transverse equivalent. Therefore, Raman beams must be precisely tuned for the interferometer device so that the interference signal can be seen. The width of the resonance line Raman is $\Delta\delta_{12} \approx 2\pi \times 0.8/\tau = 2\pi \times 0.8 v_{z0}/d$. To align horizontally, the Doppler shift $-k_{eff} v_z \sin\theta$ in the Raman transition must be less than $\Delta\delta_{12}$. That is $\theta \square\ 312\,\mu\text{rad}$ while the atomic velocity is $v_{z0} = 15\,\text{m/s}$ and the width of the Raman pulse is $d = 1\,\text{mm}$. By adjusting the distance between the lens and the fiber, and monitoring the feedback light in the fiber, you can reach $\theta \prec 91\,\mu\text{rad}$ between three Raman beams (the diameter of the fiber core is $5\,\mu\text{m}$ and, $NA = 0.11$). Otherwise, the three separated Raman beams will interact with atoms of different transverse velocities, respectively. Vertical alignment of the Raman beam is also necessary. It can be assumed that the three beams produced from the same fiber coupler are substantially parallel in the vertical direction. It can be assumed that the three beams produced from the same fiber coupler are substantially parallel in the vertical direction.

The main laser required for interferometer is a 780nm (ECDL) laser, with a typical line width of 300kHz and is stabilized using a 1.07GHz large red detuning optical meter from the $F=1 \to F'=1$ transition. Two CW diode Slave Lasers are required by Fabry-Prot GaAsP/AlGaAs. According to Fig. 10, the Slave 1 laser is injection locked to original laser. In this way, the original laser frequency is changed by an AOM that is triggered by a 60MHz signal, so the Slave 1 laser works at the $\omega_1$ frequency. Another beam of the original laser is generated by a fiber electro-optical modulator (FEOM) by an RF signal with a frequency of 6.775GHz ($\omega_1 - \omega_2 + 60\text{MHz}$) in which $\omega_1 - \omega_2$ corresponds to the hyperfine transition of $^{87}$Rb ground state. In addition to the main frequency, the FOEM output includes the $6.775 \pm 1\text{GHz}$ frequencies.

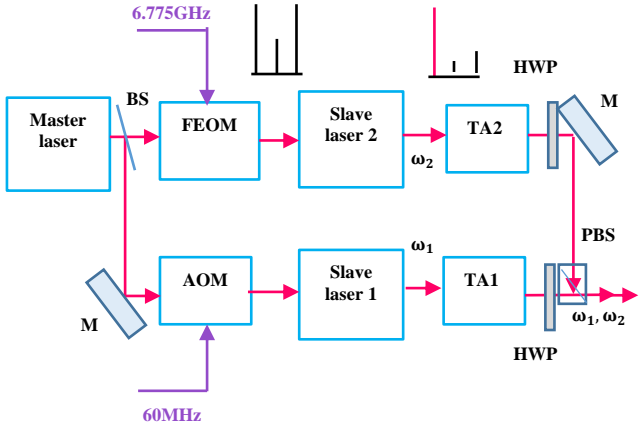

Fig 11: Laser system of interferometry section.

The -1 band is injected into the slave 2 laser. Slave 2 laser with -1 side band of FEOM output is matched with frequency-selective sideband injection-locking method. Sources of 6.775GHz and 60MHz are eventually sent to the Cs atomic clock. The width of the line of the 3dB beat note between $\omega_1$ and $\omega_2$ is 1.5Hz. Slave 1 and 2 laser beams are both amplified separately by two semiconductor laser amplifiers TA1 and TA2 to about 600mW intensities. The two outgoing beams are then combined with cross-polarization by a single mode fiber that does not change polarization and are opened by a wide-diameter 60mm Gaussian doublet fiber coupler. The Gaussian beam is blocked by a wedge with three parallel slots at a distance of $L=9.5\text{mm}$ and width $d=1\text{mm}$, allowing the intensity of two $\pi/2$ pulses to be half the pulse $\pi$ in the center of the Gaussian beam. By adjusting the distance between the lens with the focal distance of 220mm and the fiber, two $\pi/2$ Raman pulses with the same optical paths can be paralleled with the $\pi$ pulse. Three Raman beams with cross linear polarization are made of circular polarizations opposite by QWP2 wave plate. After passing through the atomic beam, the optical fields are linearized by QWP3 quarter wave plate, and then $\omega_1$ and $\omega_2$ are separated spatially by PBS. Only $\omega_1$ is reflected through the atomic beam. Propagated Raman laser beams are produced in the opposite directions for Doppler-sensitive Raman transitions as either $\sigma^+ - \sigma^+$ or $\sigma^+ - \sigma^+$. Three Raman beams are placed on the same optical table and can rotate relative to the atomic beam. By removing QWP2 and blocking the beam reflected back, Raman transitions are possible in an insensitive form to Doppler. The first cold atomic beam is observed based on interference fringes with the optical phase change of the Raman pulse, which is achieved by tilting the phase plate (9.53mm thick) which is positioned at a 45° angle in the Raman pulse and rotated with a piezoelectric transducer (PZT). To detect atomic signals, for example, interference fringes, fluorescence of $F=2$ state atoms induced by detection laser beam $(F=2 \to F'=3)$ is continuously collected by a photomultiplier tube (PMT). A fixed biased magnetic field is applied along Raman beams to hold Zeeman throughout the entire length of the interaction by a set of four wires fed by a 4A current that moves during the vacuum chamber interaction. The Mach-Zehnder atomic interferometer $\pi/2 - \pi - \pi/2$ signal can be found as follows [7]:

$$P = \frac{1}{2}\left[1 + \cos\left(\varphi_a + \varphi_\Omega + \varphi_1 - 2\varphi_2 + \varphi_3\right)\right] \quad (1)$$

where $\varphi_a = -\vec{k}_{eff} \cdot \vec{a}T^2$ is the generated phase due to acceleration and $\vec{k}_{eff} = \vec{k}_1 - \vec{k}_2$ is the effective wave vector of two lasers with $\left|\vec{k}_{eff}\right| = k_1 + k_2$ (due to the propagation of the beams in the opposite directions). $T$ is the time interval between two consecutive pulses. Therefore, by specifying this phase difference, linear acceleration can be determined which can be positioned by two integrations. The contribution of rotation in the accumulated phase difference in two paths is $\varphi_\Omega = 4\pi/(\lambda v)\vec{A}\cdot\vec{\Omega}$ in which $A$ is the area of the Sagnac loop, $\Omega$ is the rotation rate and $v$ is the atomic velocity perpendicular to the direction of the laser beam with $\lambda$ wavelength. As a result, this contribution of the phase difference determines the rotation rate, which can also be used to find direction by integrating it. $\varphi_1$, $\varphi_2$, and $\varphi_3$ are also effective phases of Raman pulses. Interference fringes are observed by pulse phase scanning by phase plate. The phase change of the interferometer output should be twice that of the $\pi$ pulse. The maximum displacement of PZT is $9\mu\text{m}$, which can cause a 30rad phase change of interference signal. By creating a proper time delay, the phase of Raman pulses can be adjusted. Then, using PZT and its frequent displacement, the detector signal can be recorded at each stage. Therefore, several equations are obtained from Eq. (1), which by solving them, the unknown phases will be determined.

Based on what was described in this paper, the first UK quantum accelerometer based on an $^{87}$Rb atom was built by a team from Imperial College London and M-Squared company [8,9]. The accelerometer works precisely, based on highly sensitive atomic interference that is used for navigation without satellites.

## 6. Summary and Conclusion

In this paper, the general principles of cold atom-based quantum navigation are briefly introduced. On of the most important properties of the atom-based navigation is non-dependency to the satellite, i.e., self-calibrating local positioning. Finally, it is worth noting that the first UK quantum accelerometer based on $^{87}$Rb cold-atoms was built by a team from Imperial College of London and M-Squared company [8,9] with accuracy is in order of $1 \times 10^{-7}$ $m/s^2$.